\documentclass[12pt,thmsa]{article}
\usepackage{sw20lart}

\input tcilatex
\begin{document}

\author{Amjad Hussain Shah Gilani \\
National Center for Physics\\
Quaid-i-Azam University\\
Islamabad 45320, Pakistan\\
Email: ahgilani@yahoo.com}
\title{Why does group theory fail to describe charge structure of particles ?}
\date{}
\maketitle

\begin{abstract}
It is pointed out that the group theory cannot describe the charge structure
of particles. Set theory is necessary to describe the charge structure of
particles but the set of charges form group.
\end{abstract}

\section{Introduction}

Despite its greater significance, the charge structure of particles has not
been adequately discussed in the existing literature. Only recently, Gilani
has pointed out a flagrant violation of group property in the charge
structure of gluons \cite{hep-ph/0404026}. This violation in the charge
structure of the gluons changes \cite{hep-ph/0410207,hep-ph/0501103} the
entire senario of the well established theory of strong interations \cite
{pt2000aug} i.e. Quantum Choromodynamics (QCD). Also, many of the
experimental results \cite{hep-ph/9706416,PLB428-383,PLB384-388} are only
confirmed by the failure of perturbative QCD \cite{PLB384-388}. An
interacting gluon model is reviewed in Ref. \cite{hep-ph/0412293}, in which
the failure of perturbative QCD is elaborated. The exploration of physics
with $b$-flavoured hadrons offers a very fertile testing ground for the
standard model (SM) description of electroweak interactions \cite
{hep-ph/0003238,PRL10-531,PTP49-652,hep-ph/9803501,hep-ph/0304132}. The
theoretical \cite{hep-ph/0105302,hep-ph/0409133} and experimental \cite
{PRL84-5283,PRL89-231801,hep-ex/0308021,hep-ex/0408138} results for the
radiative $B$-decays to kaons resonances are quite opposite. In short, the
form factors 
\begin{eqnarray*}
F_{theory}^{K^{*}} &>&F_{exp}^{K^{*}}, \\
F_{theory}^{K_1} &\ll &F_{exp}^{K_1}.
\end{eqnarray*}
Kown and Lee explained some of the possible candidates of the discrepancy 
\cite{hep-ph/0409133}.

Gilani successfully discussed the issue of charge structure in his recent
articles \cite{hep-ph/0404026,hep-ph/0410207,hep-ph/0501103}. There was also
a big confusion whether we totally escape from group concept or group
concept provide some constraints upon the choice of set of charges. The
answer of these issues are tried to discuss in this article.

\section{Charge structure by group theory}

The universe is made up of matter and it is neutral as a whole. Matter is
composed of two charges $+$ (plus) and $-$ (minus). Charge `$0$' is a
composit of charges $+$ and $-$, i.e. when we add $+$ and $-$ we obtain `$0$%
'. Let us form a group of these charges i.e. 
\[
g_1\equiv \left\{ +1,-1\right\} 
\]
The group $g_1$ is a group under multiplication but not a group under
addition. To make this a group under addition, we have to add additive
identity in it, say 
\[
g_2\equiv \left\{ +1,-1,0\right\} . 
\]
So, $g_2$ can never be group under multiplication because the inverse of `$0$%
' does not exist. We can never make $+2$ and/or $-2$ from the above groups
by using group theory.

To understand the charge structure $+2$ and/or $-2$, we have to define
another group under multiplication, i.e. 
\begin{eqnarray*}
g_3 &\equiv &\cdots ,+2,+1,+\frac 12,-\frac 12,-1,-2,\cdots \\
&=&Set\,of\,rational\,and\,irrational\,numbers
\end{eqnarray*}
but again we are unable to make `$0$' from the group $g_3$. Also, let us
take another group 
\begin{eqnarray*}
g_4 &\equiv &\cdots ,+2,+1,0,-1,-2,\cdots \\
&=&Set\,of\,whole\,numbers.
\end{eqnarray*}
The group $g_4$ is a group under addition. The groups $g_3$ and $g_4$ are
not finite groups. If we set up the charge structure with the help of groups 
$g_3$ and/or $g_4$, we are unable to get finite number of fundamental
particles.

\section{Another aspect}

The question is: Can we take any set of charges to predict the charge
structure ? Not at all. We have to take a set which form a group. But to
predict the charge structure of the particles, we have to use set theory.
Let us take an example: The square roots of unity are 
\[
S_1\equiv \left\{ +1,-1\right\} , 
\]
and the cube roots of unity are 
\[
S_2\equiv \left\{ +1,-\frac 12+i\frac{\sqrt{3}}2,-\frac 12-i\frac{\sqrt{3}}%
2\right\} . 
\]
The sets $S_1$ and $S_2$ are groups under multiplication but not groups
under addition. By using the set properties of sets $S_1$ and $S_2$, the
charge structure and prediction of particles is explained in Refs. \cite
{hep-ph/0404026,hep-ph/0410207,hep-ph/0501103}. With the help of set theory,
we can make charges like $+2$ and/or $-2$ easily and so on.

The set of charge in case of Casimir force is an empty set $\left\{
\,\,\right\} $ while in case of gravitational force, the set of charge
consists of `0' i.e. $\left\{ 0\right\} $. These sets also form group.

\section{Conclusions}

We can describe the charge structure of particles only by set theory but the
set of charges form group.

\textbf{Acknowledgements:} I am thankful to Jamil, Mariam, Shabbar for
various discussions.

\end{document}